\documentclass[12pt,preprint]{aastex}
\usepackage{epsfig}
\usepackage{natbib}

\def\gtrsim{\mathrel{\hbox{\rlap{\hbox{\lower4pt\hbox{$\sim$}}}\hbox{\raise2pt\hbox{$>$}}}}}

\newcommand\lax{${_<\atop^{\sim}}$}
\newcommand\gax{${_>\atop^{\sim}}$}

\def\lax{{$\mathrel{\hbox{\rlap{\hbox{\lower4pt\hbox{$\sim$}}}\hbox{$<$}}}$}}
\def\gax{{$\mathrel{\hbox{\rlap{\hbox{\lower4pt\hbox{$\sim$}}}\hbox{$>$}}}$}}

\begin{document}
\thispagestyle{empty}

\begin{center}
{\Large \bf The space distribution of nearby star-forming regions} \\
\end{center}

\begin{table}[h!]
\begin{tabular}{ll}
{\large \bf Authors:} & {\large \bf Laurent Loinard (UNAM)}\\
& {\large \bf Luis F.\ Rodr\'{\i}guez (UNAM)}\\
&{\large \bf Amy J.\ Mioduszewski (NRAO)}\\
\end{tabular}
\end{table}

\begin{table}[h!]
\begin{tabular}{ll}
{\large Contact Author:}&{\large \bf Laurent Loinard} \\
&{\large Centro de Radiostronom\'{\i}a y Astrof\'{\i}sica,} \\
&{\large Universidad Nacional Aut\'onoma de M\'exico,}\\
&{\large Morelia , Mexico}\\
&{\large e-mail: l.loinard@astrosmo.unam.mx} \\
& \\
{\Large Frontier Area 1:}&{\Large The Planetary Systems and Star Formation} \\
\label{tab:contact}
\end{tabular}
\end{table}
\parindent 0pt

{\large {\bf Goal: Determine the distance, structure and dynamics of all northern star-forming region within 1 kpc of the Sun}}
\parskip 0pt

\bigskip

\begin{itemize}
\itemsep 0pt

\item{\large{\it What is the local space distribution of star-formation in the Solar neighborhood?}}

\item{\large{\it Is the internal structure and dynamics of nearby star-forming regions in agreement with theoretical models?}}

\item{\large{\it What are the precise intrinsic properties of low-mass young stars?}}

\item{\large{\it How do those properties compare with theoretical expectations?}}

\end{itemize}

\parindent 0pt

\clearpage
\setcounter{page}{1}

\begin{abstract}

Multi-epoch radio-interferometric observations of young
stellar objects can be used to measure their displacement 
over the celestial sphere with a level of accuracy that currently 
cannot be attained at any other wavelength. In particular, the 
accuracy achieved using carefully calibrated, phase-referenced 
observations with Very Long Baseline Interferometers such as 
NRAO's  {\em Very Long Baseline Array} is better 
than 50 micro-arcseconds. This is sufficient to measure the trigonometric
parallax and the proper motion of any radio-emitting young star within
several hundred parsecs of the Sun with an accuracy better than a few
percent. Using that technique, the mean distances to Taurus, Ophiuchus, 
Perseus and Orion have already been measured to unprecedented  
accuracy. 

With improved telescopes and equipment, the distance to all 
star-forming regions within 1 kpc of the Sun and beyond, as well as their
internal structure and dynamics could be determined. This would
significantly improve our ability to compare the observational properties
of young stellar objects with theoretical predictions, and would
have a major impact on our understanding of low-mass star-formation.

\end{abstract}

\section{Introduction}

Astrometric observations of young stellar objects can provide a wealth
of important information on their properties. First and foremost, an
accurate trigonometric parallax measurement is a pre-requisite to the
derivation, from observational data, of their most important
characteristics (luminosity, age, mass, etc.). Unfortunately, even in
the current post-Hipparcos era, the distance to even the nearest
star-forming regions (Taurus, Ophiuchus, Perseus, etc.) is rarely
known to better than 20 to 30\% (e.g.\ Knude \& H\"og 1998; Bertout et
al.\ 1999). At this level of accuracy, the mass of a binary system
derived from observations of its orbital motion would be uncertain by
a factor of two. This unsatisfactory state of affairs is largely the
result of the fact that young stars are still embedded in their opaque
parental cloud. They are, therefore, dim in the visible bands that
were observed by Hipparcos. The proper motions that can be derived 
from astrometric observations of young stars are also of interest,
particularly to study the internal dynamics of star-forming regions.

Since observations of young stars in the visible range is limited by
the effect of dust extinction, one must turn to a more favorable
wavelength regime in order to obtain high quality astrometric
data. Radio observations, particularly using large interferometers is
currently the best prospect because (i) the interstellar medium is
largely transparent at these wavelengths, and (ii) the astrometry
delivered by radio-interferometers is extremely accurate and
calibrated against fixed distant quasars. Of course, only those young
stars associated with radio sources are potential targets. This is
currently a significant limiting factor, which restricts the number of 
accessible targets to a few handfuls. With a relatively modest  
increase (a factor of 10) in the sensitivity of Very Long Baseline 
Interferometers, however, hundreds of young stars would become
adequate targets.

\begin{figure*}[!t]
\centerline{\includegraphics[width=0.39\columnwidth,angle=270]{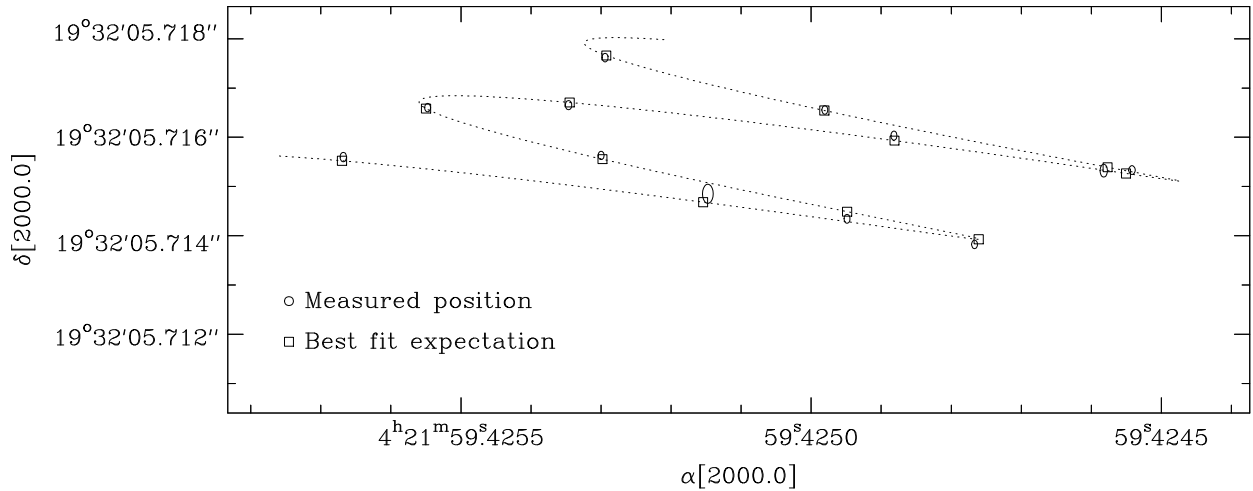}}
  \caption{Measured positions of T Tau Sb and best fit combining 
  proper motion and trigonometric parallax (from Loinard et al.\ 2007). 
  The observed positions are shown as ellipses, the size of which 
  represents the magnitude of the errors.}  
\end{figure*}

\section{The distance to nearby star-forming regions}

Using multi-epoch observations obtained using Very Long
Baseline Interferometry (VLBI), the mean distance to several
nearby star-forming regions have already been measured
with unprecedented accuracy. We will briefly describe these
results here to demonstrate the potential of VLBI instruments.

\subsection{Taurus}

Until now, the distances to 4 young stars embedded in the 
Taurus complex have been measured using multi-epoch Very 
Long Baseline Array (VLBA) observations (see Loinard et al.\ 
2005, 2007, Torres et al.\ 2007, 2009). As an example, the case of
T Tau Sb (one of the southern companions of the famous
young star T Tauri) is shown in Fig.\ 1. The trigonometric parallax 
obtained from these observations is 6.82 $\pm$ 0.03 mas, 
corresponding to a distance of 146.7 $\pm$ 0.6 pc (Loinard et 
al.\ 2005, 2007). Similar results were obtained for the other three 
sources, resulting in a mean distance of about 142 pc for the entire 
Taurus complex. This is in good agreement with the value of 140 
$\pm$ 15 pc traditionally used for Taurus (e.g.\ Kenyon et al.\ 1994, 
Bertout et al.\ 1999).

Perhaps more interesting than that mean distance, however, 
is the information on the three-dimensional structure of Taurus
revealed by our observations. The extent of Taurus on the plane 
of the sky is about 10$^\circ$ (25 pc), and one would expect
its depth to be similar. Our observations confirm this (Fig.\ 2) since 
HP Tau, to the east of the complex is at 161.2 $\pm$ 0.9 pc, about 
30 pc farther than Hubble 4 or HDE~283572 in the central region
(the mean distance of Hubble 4 and HDE~283572 is 130.7 $\pm$ 0.6 pc). 
T Tau appears to be at an intermediate distance. It is important to
realize that because of this significant depth, even if the mean
distance of the Taurus association were known to infinite accuracy, we
could still make errors as large as 10--20\% by using the mean
distance indiscriminately for all sources in Taurus. To reduce this
systematic source of error, one needs to establish the
three-dimensional structure of the Taurus association, and
our existing observations represent the first step in that
direction.

In addition to their trigonometric parallaxes, the VLBA astrometry
of the several stars in Taurus provides an accurate determination
of their proper motions. Combining this information with existing 
radial velocity measurements, we can reconstruct the full velocity
vector of each star. As Fig.\ 2 shows, there is a systematic 
difference in the orientation and amplitude of the tangential 
velocity between the central and the eastern portions of Taurus.
This echoes the situation with radial velocities which systematically
differ by about 3 km s$^{-1}$ between the western and the eastern 
edge of the complex. Clearly, if data similar to those presented here
were available for tens of stars, it would become possible to 
study in detail the internal dynamics of the Taurus complex.

\begin{figure*}[!t]
\centerline{\includegraphics[width=0.7\columnwidth,angle=270]{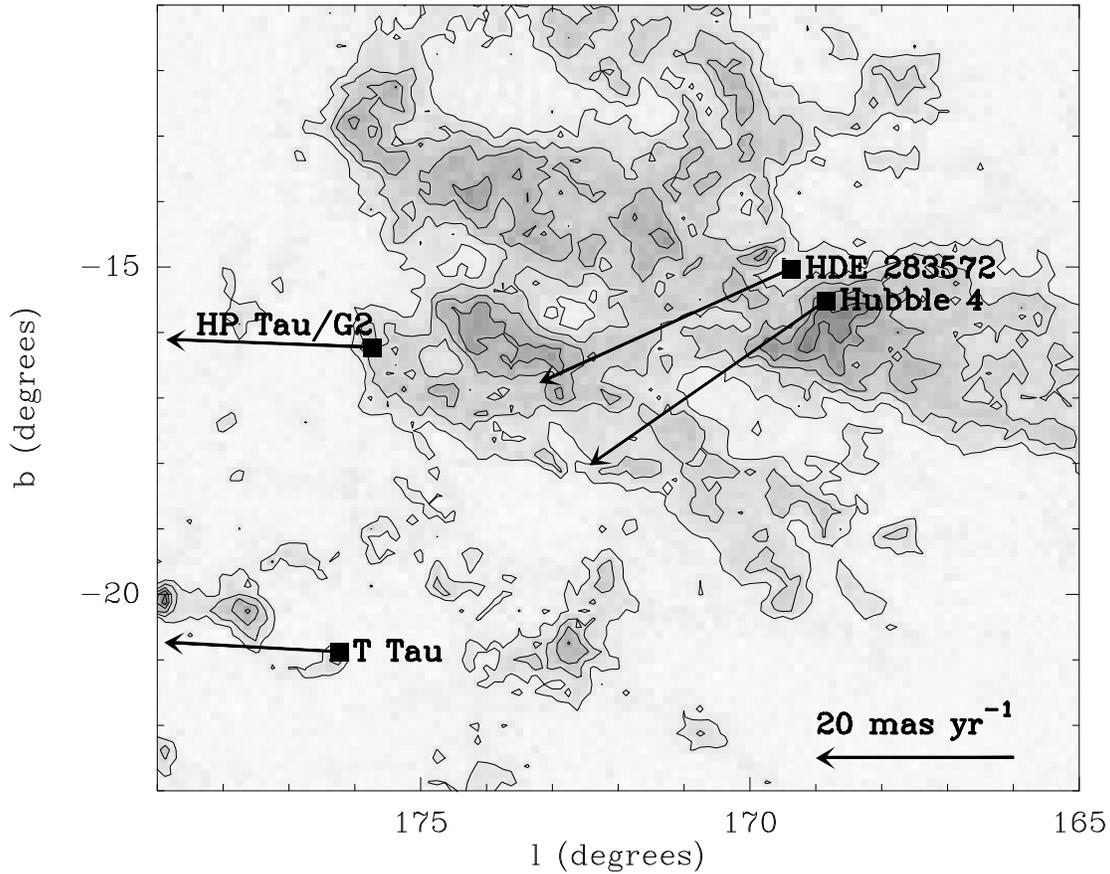}}
  \caption{Position and tangential velocity vectors of the four young
  stars in Taurus considered here, superimposed onto the CO(1-0) map of
  Taurus from Dame et al.\ (2001).}  
\end{figure*}

\subsection{The Ophiuchus core}

Ophiuchus is, together with Taurus, one the best studied regions of
low-mass star-formation. It has long been thought to be at 165 pc (Chini 1981), but
somewhat shorter distances (120--135 pc) have recently been proposed
(Knude \& H\"og 1998; Mamajek et al.\ 2008).

Two young stars embedded within Ophiuchus were observed with 
the VLBA at 6 and 7 epochs, respectively (Loinard et al.\ 2008).
The weighted mean of their parallaxes is 8.33 $\pm$ 0.30, 
corresponding to a distance of 120.0$^{+4.5}_{-4.3}$ pc for
the Ophiuchus complex. This is in good agreement with the value
proposed by Knude \& H\"og (1998) and, more recently, by Lombardi et
al.\ (2008). Note that this determination of the distance to
Ophiuchus is accurate to better than 4\% ($\equiv$ 5 pc). This is to
be compared to the situation prior to our observations, when the
uncertainty was between 120, 140 or 160 pc. It should be noticed, 
however, that this distance is that of the Ophiuchus core. Other 
parts of the Ophiuchus (especially the so-called streamers) could
be at a somewhat different distance. Also, we observed a third
star in the direction of Ophiuchus, and  found to be at about 
170 pc. If confirmed, this result could indicate the presence of
several (unrelated) regions of star-formation along the line of
sight.

\subsection{Perseus and Orion}

The distance to NGC 1333 within the Perseus region was
measured using the Japanese VLBI system (called VERA)
by Hirota et al.\ (2008). Rather than using continuum radio 
emission from the active magnetospheres of young stars, 
Hirota et al.\ used water masers associated with the outflow 
powered by SVS13. They obtain a distance of 235 $\pm$
18 pc. It is still not completely clear if this distance should 
be used for the entire Perseus complex or if other regions
with Perseus (e.g.\ IC 348) might be somewhat farther.

The distance to the Orion Nebula was measured by Menten et 
al.\ (2007) using VLBA observations of 4 embedded stars.
They obtain 414 $\pm$ 7 pc. As in the case of Ophiuchus or Perseus, it
is not clear if this value should be used for the entire Orion
complex or if regions far from the nebula might be at somewhat
different distances.

\section{Future possibilities and instrumentation needs}

The observations presented above have already improved
significantly our knowledge of the distribution of star-forming 
regions around the Sun. Yet, they merely represent the ``tip 
of the iceberg'' of what could be done in the coming decades
with improved VLBI systems. 

The sensitivity of existing VLBI instruments, and of the VLBA 
in particular, is limited. As a consequence, observations such 
as those described above can only be obtained for relatively 
bright sources, and even then, they require fairly large amounts 
of  telescope time (over 500 hours of VLBA time were necessary 
to collect the Taurus and Ophiuchus data presented in \S 2). 
The luminosity function of radio sources associated with
young stars is still poorly constrained, but it is clear that
increasing the sensitivity of the VLBA by one order of 
magnitude would allow similar observations to be obtained for
hundreds of stars instead of a handful of them. 

This would
allow the determination of the distance to all northern star-forming 
regions within 1 kpc of the Sun. By including tens of stars in
each of these regions, it would also become possible to establish
their structure and internal dynamics. This would, in particular,
allow a direct comparison between the observed dynamics of 
star-forming regions, and the predictions of theoretical models.
A direct  product of the observations described above would be 
the construction of an unprecedented sample of young stellar 
sources with extremely well-measured intrinsic properties. This
would allow a very detailed comparison with theoretical model of 
early stellar evolution. In summary, the proposed observations would 
allow us to answer the following four important questions:

\begin{itemize}

\item{{\bf What is the local space distribution of star-formation in the Solar neighborhood?}}

\item{{\bf Is the internal structure and dynamics of nearby star-forming regions in agreement with theoretical models?}}

\item{{\bf What are the precise intrinsic properties of low-mass young stars?}}

\item{{\bf How do those properties compare with theoretical expectations?}}

\end{itemize}

Of course, the answer to these question would have important
consequences for various related issues, such as the determination
of the low-mass end of the IMF, or the eventual formation of planets 
around low-mass stars.

It is interesting to note that the sensitivity increase necessary to
carry out much this program does not require new technology. The
sensitivity of the VLBA could be improved by one order
of magnitude by increasing the frequency bandwidth 
recorded (from 32 MHz currently to 2 GHz). Besides increasing 
enormously the number of
stars that could be studied, such a significant increase in sensitivity 
would also improve the astrometry for each target. The reason
for this is the following. The quality of the astrometry reached in 
a VLBI observation depends directly on the distance between the
target and the calibrating quasar. An increase in sensitivity 
would make many weak quasars suitable calibrators, and
would, therefore, reduce dramatically the distance between
any target and the nearest available calibrator. The gain would
be very significant: a VLBA one to tow orders of magnitude more sensitive 
could routinely deliver micro-arcsecond astrometry! 

If the above improvements in recording capabilities were
combined with a significant increase in collecting area, observations
similar to those presented earlier could be extended to much
larger distances,  and it would become possible to effectively 
map out the Milky Way (recall that the trigonometric parallax of a 
source at 20 kpc is 50 micro-arcseconds). Note, indeed, that the distances to
several massive star-forming regions along nearby spiral arms
(at a few kpc) have already been measured to remarkable precision using
the VLBA (the targets in these cases were water masers, see
Xu et al.\ 2006). The necessary increase in collecting area could 
be achieved as part of the SKA-High project, if some of the groups 
of antennas for that project were built near existing VLBA 
antennas, or between the current VLBA antennas and the EVLA site.

Finally, we would like to point out that the improvements
proposed here would fit naturally within the framework of the 
{\em North America Array} initiative (http://www.nrao.edu/nio/naa/)
coordinated by Jim Ulvestad and submitted to the decadal survey.

\section{Conclusions}

In this paper, we showed that multi-epoch VLBI observations
young stars allows the determination of their trigonometric 
parallaxes with accuracies of a few percent. In the next 
decade or two, it would become possible to  establish the 
distribution of star formation in the Solar Neighborhood
($d$ $<$ 1 kpc) or farther using improved VLBI systems.
This would have a major impact on our understanding of
star-formation because it would allow us to compare the
both the structure and dynamics of star-forming regions
and their intrinsic properties with theoretical expectations.

\end{document}